\documentclass[a4paper,11pt]{article}
\usepackage{pos}
\usepackage[utf8]{inputenc}
\usepackage{bm}
\usepackage{graphicx}
\usepackage{wrapfig}

\title{
Lambda-Nucleon and Sigma-Nucleon potentials from space-time correlation function on the lattice
}
\ShortTitle{
$\Lambda N$ and $\Sigma N$ potentials from space-time correlation function on the lattice
}

\author*[a,b]{Hidekatsu Nemura 
}

\affiliation[a]{Research Center for Nuclear Physics, Osaka University, 
\\
  10, Mihogaoka, Ibaraki-shi, Osaka 567-0047, Japan
}

\affiliation[b]{Yukawa Institute for Theoretical Physics, Kyoto University, 
\\
Kitashirakawa Oiwakecho, Sakyo-ku, Kyoto 606-8502, Japan
}

\emailAdd{hidekatsu.nemura@rcnp.osaka-u.ac.jp 
}

\abstract{
 The hyperon-nucleon interaction with the strangeness $S=-1$ region is complicated and difficult to investigate because 
 its flavor sector involves all the irreducible representation except the flavor singlet and has the worst signal-to-noise 
 ratio among the strangeness regions. 
 In order to overcome such difficulties the content of this report is twofold: 
(i) 
We present an implementation of extended effective baryon block algorithm. 
This is a straightforward extension of the original which was reported in LATTICE 2013. 
(ii) We perform single channel analysis for the $\Lambda N$ system at nearly physical 
quark masses corresponding to 
$(m_\pi,m_K)\approx(146,525)$~MeV and large volume 
 $(La)^4=(96a)^4\approx$ (8.1 fm)$^4$. 
Scattering phase shifts for $\Lambda N$ system are presented. 
}

\FullConference{%
 The 38th International Symposium on Lattice Field Theory, LATTICE2021
  26th-30th July, 2021
  Zoom/Gather@Massachusetts Institute of Technology
}

\begin{document}
\maketitle


\section{Introduction}

Atomic nucleus is a finite quantum many-body system of nucleons bound by the strong interaction (nuclear force). 
In the low energy region, nuclear phenomena have been studied with much success by assuming that 
the nucleon is the fundamental degree of freedom. 
On the other hand, %
in systems including strangeness, the hyperonic nuclear forces have still large ambiguities 
because of the lack of sufficient experimental data. 
Neutron stars with twice the solar mass have been 
observed\cite{Demorest:2010bx,Antoniadis:2013pzd,Fonseca:2021wxt}, 
and an accurate understanding of the hyperonic nuclear 
force is a major issue in 
understanding the structure of dense nuclear states such as those near the core of neutron stars.

Comprehensive study of generalized baryon-baryon ($BB$) interaction with containing strangeness is one of 
the important subject. %
HAL QCD method\cite{Aoki:2012tk} is one of the fascinating approaches. 
In order to calculate a wide range of %
$BB$ interactions %
simultaneously, we had presented 
the effective baryon block algorithm in LATTICE 2013\cite{Nemura:2014eta}. 
Since this algorithm does not impose any restrictions on the quark fields 
on each baryon in the source, there is no need for each quark field 
in the source to be spatially identical between the baryons. 
The Wick contraction can be performed appropriately no matter what 
quantum state is considered.

The 
coupled-channel potentials of $\Lambda N-\Sigma N$ system 
at nearly physical quark masses corresponding to $(m_{\pi}, m_{K}) = (146, 525)$~MeV 
with large volume $(L a)^{4} = (96 a)^{4} = (8.1~\mathrm{fm})^{4}$ 
had been %
reported in LATTICE 2017\cite{Nemura:2017vjc}. 
The $\Sigma N-\Sigma N$ potential in the $^1S_0$ channel has large statistical noise 
which impedes further analysis. %

The purpose of this report is twofold: 
(i) We present an implementation of extended effective baryon block algorithm. 
It is straightforward from the first implementation~\cite{Nemura:2015yha} 
since the original algorithm does not impose any restrictions on the quark fields 
on each baryon in the source. 
(ii) We present single channel analysis for the $\Lambda N$ system 
at nearly physical quark masses corresponding to $(m_{\pi}, m_{K}) = (146, 525)$~MeV 
with large volume $(L a)^{4} = (96 a)^{4} = (8.1~\mathrm{fm})^{4}$. 
By employing the single-channel analysis the statistical uncertainty of 
the single channel $\Lambda N$ potential is reduced. 
We can derive the scattering phase shifts below the $\Sigma N$ threshold 
by employing an analytical functional form for representing the lattice (discretized) 
potential.

\section{%
HAL QCD method 
with 
effective baryon block algorithm
}

\begin{figure}[t]
  \centering \leavevmode 
  \includegraphics[width=0.989\textwidth,angle=0]{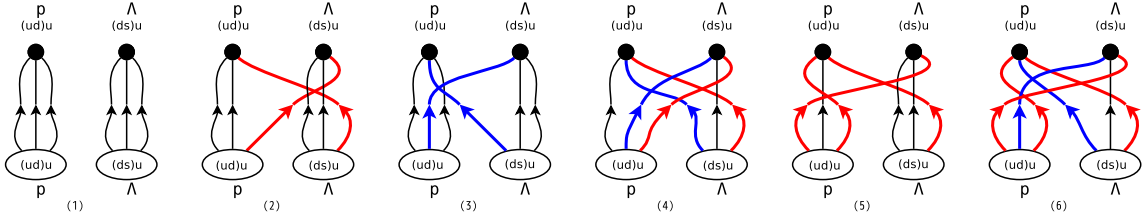}
  \footnotesize
 \caption{Diagrammatic representation of the four-point correlation function
   $\langle p\Lambda\overline{p\Lambda_{\mathrm{dsu}}}\rangle$. 
   The interpolating field $\Lambda_{\mathrm{dsu}}$ is used as a representative of $\Lambda$. 
 \label{Fig_FFT_Lp}}
\end{figure}

The baseline quantity for studying the $\Lambda N$ interaction with the HAL QCD method 
is the four-point correlation function (4pt-correlator) in the center-of-mass frame, 
\begin{equation}
 { F}_{\alpha_{1}\alpha_{2},\alpha_{3}\alpha_{4}}
     ^{\langle p\Lambda\overline{p\Lambda}\rangle}(\vec{r},t-t_0) = 
 \sum_{\vec{X}}
 \left\langle 0
  \left|
   p_{{\alpha_1}}(\vec{X}+\vec{r},t)
   \Lambda_{{\alpha_2}}(\vec{X},t)
   \overline{{\cal J}_{p_{\alpha_{3}} \Lambda_{\alpha_{4}}}
                     (t_0)}
  \right| 0 
 \right\rangle. 
\end{equation}
The $p$ and $\Lambda$ denote the 
interpolating fields of proton and $\Lambda$ 
which comprise up, down and strange quark fields, $u$, $d$, $s$, 
\begin{equation}
 \!\!\!
 \begin{array}{llll}
  p \! = \! X_{\mathrm{udu}}, \!
  &
  \Lambda \! = \! {1\over \sqrt{6}} \left( X_{\mathrm{dsu}} \! + \! X_{\mathrm{sud}} \! - \! 2 X_{\mathrm{uds}} \right),\!
  &
  \mathrm{with} ~ X_{\mathrm{fgh}} \! = \! \varepsilon_{abc} \! \left( \mathrm{f}_a C\gamma_5 \mathrm{g}_b \right) \! \mathrm{h}_c, 
  &
  (\mathrm{f},\mathrm{g},\mathrm{h}) \! \in \! \left\{ u, d, s \right\}. 
 \end{array}
 \label{BaryonOperatorsProtonLambda}
\end{equation}
%
The 4pt-correlator is evaluated through the effective baryon block 
algorithm~\cite{Nemura:2015yha,Nemura:2016sty} 
together with Fast Fourier Transform (FFT). 
For simplicity, we show only the contributions from $\overline{X}_{\mathrm{dsu}}$ 
in the $\overline{\Lambda}$ in the source. 
%
\begin{eqnarray}
 && {F}^{\langle p \Lambda\overline{pX_{\mathrm{dsu}}}\rangle}
    _{{\alpha_1}{\alpha_2},{\alpha_3}{\alpha_4}}(\vec{r})
 =
 \sum_{\vec{X}}
 \left\langle 0
  \left|
   p_{{\alpha_1}}(\vec{X}+\vec{r},t)
   \Lambda_{{\alpha_2}}(\vec{X},t)
   \overline{{\cal J}_{p_{\alpha_{3}} X_{\mathrm{dsu},\alpha_{4}}}
                     (t_0)}
  \right| 0 
 \right\rangle,
 \nonumber \\
 &=& 
  {1\over L^3}
   \sum_{\vec{q}}
    \left(
%
    [\widetilde{      p}_{\alpha_1\alpha_3}^{(1)}]( \vec{q}) 
    [\widetilde{\Lambda}_{\alpha_2\alpha_4}^{(1)}](-\vec{q})
    -
    [\widetilde{      p}_{\alpha_1\alpha_4}^{(2)}]_{c_{3}^{\prime},c_{6}^{\prime}}( \vec{q}) 
    [\widetilde{\Lambda}_{\alpha_2\alpha_3}^{(2)}]_{c_{3}^{\prime},c_{6}^{\prime}}(-\vec{q})
    \right.
    \nonumber
    \\
    && %
    \left.
    -
    [\widetilde{      p}_{\alpha_1\alpha_3}^{(3)}]_{c_{2}^{\prime},\alpha_{2}^{\prime},c_{4}^{\prime},\alpha_{4}^{\prime}}( \vec{q}) 
    [\widetilde{\Lambda}_{\alpha_2\alpha_4}^{(3)}]_{c_{2}^{\prime},\alpha_{2}^{\prime},c_{4}^{\prime},\alpha_{4}^{\prime}}(-\vec{q})
    \right.
    \left.
    +
    [\widetilde{      p}_{\alpha_1\alpha_4}^{(4)}]_{c_{1}^{\prime},\alpha_{1}^{\prime},c_{5}^{\prime},\alpha_{5}^{\prime}}( \vec{q}) 
    [\widetilde{\Lambda}_{\alpha_2\alpha_3}^{(4)}]_{c_{1}^{\prime},\alpha_{1}^{\prime},c_{5}^{\prime},\alpha_{5}^{\prime}}(-\vec{q})
    \right.
    \nonumber
    \\
    && %
    \left.
    +
    [\widetilde{      p}_{\alpha_1\alpha_3\alpha_4}^{(5)}]_{c_{1}^{\prime},\alpha_{1}^{\prime},c_{6}^{\prime}}( \vec{q}) 
    [\widetilde{\Lambda}_{\alpha_2                }^{(5)}]_{c_{1}^{\prime},\alpha_{1}^{\prime},c_{6}^{\prime}}(-\vec{q})
    \right.
    \left.
    -
    [\widetilde{      p}_{\alpha_1\alpha_3\alpha_4}^{(6)}]_{c_{3}^{\prime},c_{5}^{\prime},\alpha_{5}^{\prime}}( \vec{q}) 
    [\widetilde{\Lambda}_{\alpha_2                }^{(6)}]_{c_{3}^{\prime},c_{5}^{\prime},\alpha_{5}^{\prime}}(-\vec{q})
    \right)
    {\rm e}^{i\vec{q}\cdot\vec{r}},
  \label{pL.pLXup.FFT}
\end{eqnarray}
where
\begin{eqnarray}
 &&
  \begin{array}{l}
   ~[\widetilde{      p}_{\alpha_1\alpha_3}^{(1)}]( \vec{q}) 
    =
    [\widetilde{      p}_{\alpha_{1}}^{(0)}](\vec{q};~\bm{\xi}_{{1}{2}{3}}^\prime) 
    \varepsilon_{c_{{1}}^{\prime} c_{{2}}^{\prime} c_{{3}}^{\prime}} (C\gamma_5)_{\alpha_{1}^{\prime}\alpha_{2}^{\prime}} \delta_{\alpha_{3}^{\prime}\alpha_{3}},
   \\
   ~[\widetilde{\Lambda}_{\alpha_2\alpha_4}^{(1)}](-\vec{q})
   =
   [\widetilde{\Lambda}_{\alpha_{2}}^{(0)}](-\vec{q};\bm{\xi}_{{4}{5}{6}}^\prime) 
   \varepsilon_{c_{{4}}^{\prime} c_{{5}}^{\prime} c_{{6}}^{\prime}} (C\gamma_5)_{\alpha_{4}^{\prime}\alpha_{5}^{\prime}} \delta_{\alpha_{6}^{\prime}\alpha_{4}},
  \end{array}
  \\
 &&
  \begin{array}{l}
   ~[\widetilde{      p}_{\alpha_1\alpha_4}^{(2)}]_{c_{3}^{\prime}c_{6}^{\prime}}( \vec{q}) 
    =
    [\widetilde{      p}_{\alpha_{1}}^{(0)}](\vec{q};~\bm{\xi}_{{1}{2}{6}}^\prime) 
    \varepsilon_{c_{{1}}^{\prime} c_{{2}}^{\prime} c_{{3}}^{\prime}} (C\gamma_5)_{\alpha_{1}^{\prime}\alpha_{2}^{\prime}} \delta_{\alpha_{6}^{\prime}\alpha_{4}},
   \\
   ~[\widetilde{\Lambda}_{\alpha_2\alpha_3}^{(2)}]_{c_{3}^{\prime}c_{6}^{\prime}}(-\vec{q})
   =
   [\widetilde{\Lambda}_{\alpha_{2}}^{(0)}](-\vec{q};\bm{\xi}_{{4}{5}{3}}^\prime) 
   \varepsilon_{c_{{4}}^{\prime} c_{{5}}^{\prime} c_{{6}}^{\prime}} (C\gamma_5)_{\alpha_{4}^{\prime}\alpha_{5}^{\prime}} \delta_{\alpha_{3}^{\prime}\alpha_{3}},
  \end{array}
  \\
 &&
  \begin{array}{l}
   ~[\widetilde{      p}_{\alpha_1\alpha_3}^{(3)}]_{c_{2}^{\prime}\alpha_{2}^{\prime}c_{4}^{\prime}\alpha_{4}^{\prime}}( \vec{q}) 
    =
    [\widetilde{      p}_{\alpha_{1}}^{(0)}](\vec{q};~\bm{\xi}_{{1}{4}{3}}^\prime) 
    \varepsilon_{c_{{1}}^{\prime} c_{{2}}^{\prime} c_{{3}}^{\prime}} (C\gamma_5)_{\alpha_{1}^{\prime}\alpha_{2}^{\prime}} \delta_{\alpha_{3}^{\prime}\alpha_{3}},
   \\
   ~[\widetilde{\Lambda}_{\alpha_2\alpha_4}^{(3)}]_{c_{2}^{\prime}\alpha_{2}^{\prime}c_{4}^{\prime}\alpha_{4}^{\prime}}(-\vec{q})
   =
   [\widetilde{\Lambda}_{\alpha_{2}}^{(0)}](-\vec{q};\bm{\xi}_{{2}{5}{6}}^\prime) 
   \varepsilon_{c_{{4}}^{\prime} c_{{5}}^{\prime} c_{{6}}^{\prime}} (C\gamma_5)_{\alpha_{4}^{\prime}\alpha_{5}^{\prime}} \delta_{\alpha_{6}^{\prime}\alpha_{4}},
  \end{array}
  \\
 &&
  \begin{array}{l}
   ~[\widetilde{      p}_{\alpha_1\alpha_4}^{(4)}]_{c_{1}^{\prime}\alpha_{1}^{\prime}c_{5}^{\prime}\alpha_{5}^{\prime}}( \vec{q}) 
    =
    [\widetilde{      p}_{\alpha_{1}}^{(0)}](\vec{q};~\bm{\xi}_{{1}{4}{6}}^\prime) 
    \varepsilon_{c_{{4}}^{\prime} c_{{5}}^{\prime} c_{{6}}^{\prime}} (C\gamma_5)_{\alpha_{4}^{\prime}\alpha_{5}^{\prime}} \delta_{\alpha_{6}^{\prime}\alpha_{4}},
   \\
   ~[\widetilde{\Lambda}_{\alpha_2\alpha_3}^{(4)}]_{c_{1}^{\prime}\alpha_{1}^{\prime}c_{5}^{\prime}\alpha_{5}^{\prime}}(-\vec{q})
   =
   [\widetilde{\Lambda}_{\alpha_{2}}^{(0)}](-\vec{q};\bm{\xi}_{{2}{5}{3}}^\prime) 
   \varepsilon_{c_{{1}}^{\prime} c_{{2}}^{\prime} c_{{3}}^{\prime}} (C\gamma_5)_{\alpha_{1}^{\prime}\alpha_{2}^{\prime}} \delta_{\alpha_{3}^{\prime}\alpha_{3}},
  \end{array}
  \\
 &&
  \begin{array}{l}
   ~[\widetilde{      p}_{\alpha_1\alpha_3\alpha_4}^{(5)}]_{c_{1}^{\prime}\alpha_{1}^{\prime}c_{6}^{\prime}}( \vec{q}) 
    =
    [\widetilde{      p}_{\alpha_{1}}^{(0)}](\vec{q};~\bm{\xi}_{{3}{2}{6}}^\prime) 
    \varepsilon_{c_{{1}}^{\prime} c_{{2}}^{\prime} c_{{3}}^{\prime}} (C\gamma_5)_{\alpha_{1}^{\prime}\alpha_{2}^{\prime}} \delta_{\alpha_{3}^{\prime}\alpha_{3}}
                                                                                                                          \delta_{\alpha_{6}^{\prime}\alpha_{4}},
   \\
   ~[\widetilde{\Lambda}_{\alpha_2                }^{(5)}]_{c_{1}^{\prime}\alpha_{1}^{\prime}c_{6}^{\prime}}(-\vec{q})
    =
    [\widetilde{\Lambda}_{\alpha_{2}}^{(0)}](-\vec{q};\bm{\xi}_{{4}{5}{1}}^\prime) 
    \varepsilon_{c_{{4}}^{\prime} c_{{5}}^{\prime} c_{{6}}^{\prime}} (C\gamma_5)_{\alpha_{4}^{\prime}\alpha_{5}^{\prime}},
  \end{array}
  \\
 &&
  \begin{array}{l}
   ~[\widetilde{      p}_{\alpha_1\alpha_3\alpha_4}^{(6)}]_{c_{3}^{\prime}c_{5}^{\prime}\alpha_{5}^{\prime}}( \vec{q}) 
    =
    [\widetilde{      p}_{\alpha_{1}}^{(0)}](\vec{q};~\bm{\xi}_{{3}{4}{6}}^\prime) 
    \varepsilon_{c_{{4}}^{\prime} c_{{5}}^{\prime} c_{{6}}^{\prime}} (C\gamma_5)_{\alpha_{4}^{\prime}\alpha_{5}^{\prime}}
                                                                                                                         \delta_{\alpha_{3}^{\prime}\alpha_{3}}
                                                                                                                         \delta_{\alpha_{6}^{\prime}\alpha_{4}},
   \\
   ~[\widetilde{\Lambda}_{\alpha_2                }^{(6)}]_{c_{3}^{\prime}c_{5}^{\prime}\alpha_{5}^{\prime}}(-\vec{q})
    =
    [\widetilde{\Lambda}_{\alpha_{2}}^{(0)}](-\vec{q};\bm{\xi}_{{2}{5}{1}}^\prime) 
    \varepsilon_{c_{{1}}^{\prime} c_{{2}}^{\prime} c_{{3}}^{\prime}} (C\gamma_5)_{\alpha_{1}^{\prime}\alpha_{2}^{\prime}},
  \end{array}
  \label{EffectiveBaryonBlocksProtonLambdaStructuredForm}
\end{eqnarray}
%
\begin{eqnarray}
  ~[p_{\alpha_{1}}^{(0)}](\vec{x};~\bm{\xi}_{{1}{2}{3}}^\prime) 
 \!\!&=&\!\!
 \varepsilon_{b_{{1}} b_{{2}} b_{{3}}} (C\gamma_5)_{\beta_{1}\beta_{2}} \delta_{\beta_{3}\alpha_{1}}
 \det\left|
  \begin{array}{cc}
    \langle u(\zeta_{1}) \bar{u}(\xi_{1}^{\prime}) \rangle &
    \langle u(\zeta_{1}) \bar{u}(\xi_{3}^{\prime}) \rangle
    \\
    \langle u(\zeta_{3}) \bar{u}(\xi_{1}^{\prime}) \rangle &
    \langle u(\zeta_{3}) \bar{u}(\xi_{3}^{\prime}) \rangle
  \end{array}
 \right|
 \langle d(\zeta_{2}) \bar{d}(\xi_{2}^{\prime}) \rangle,
 \\
 ~
 [\Lambda_{\alpha_{2}}^{(0)}](\vec{y};~\bm{\xi}_{{4}{5}{6}}^\prime) 
 \!\!&=&\!\!
 {1\over \sqrt{6}}
 \varepsilon_{b_{4} b_{5} b_{6}}
  \left\{
   (C\gamma_5)_{\beta_{4}\beta_{5}} \delta_{\beta_{6}\alpha_{2}}
   + (C\gamma_5)_{\beta_{5}\beta_{6}} \delta_{\beta_{4}\alpha_{2}}
   -2 (C\gamma_5)_{\beta_{6}\beta_{4}} \delta_{\beta_{5}\alpha_{2}}
  \right\}
  \nonumber
  \\
  &&
  \qquad
  \times
     {\langle u(\zeta_{6}) \bar{u}(\xi_{6}^{\prime}) \rangle}
      \langle d(\zeta_{4}) \bar{d}(\xi_{4}^{\prime}) \rangle
      \langle s(\zeta_{5}) \bar{s}(\xi_{5}^{\prime}) \rangle.
\end{eqnarray}
%
%
Six terms on right-hand side in Eq.~(\ref{pL.pLXup.FFT}) correspond to 
six diagrams in the Fig.~\ref{Fig_FFT_Lp}. 
The subscripts $c^{\prime}$ ($\alpha^{\prime}$) are for color 
(Dirac spinor) that run from 1 to $N_{c}=3$ ($N_{\alpha}=4$). 
We have introduced shorthand notation, 
${\bm{\xi}_{123}^{\prime}}=\left( \xi_{1}^{\prime}, \xi_{2}^{\prime}, \xi_{3}^{\prime} \right)$, 
and each $\xi_{i}^{\prime}$ ($\zeta_{i}$) is the spin-color-space-time 
coordinate of the quark field on the source (sink) side. 
All of the contributions from $\overline{X}_{\mathrm{dsu}}$, 
$\overline{X}_{\mathrm{sud}}$, and $\overline{X}_{\mathrm{uds}}$ in 
$\overline{\Lambda}$ are taken into account in the actual computation. 
By employing the effective block algorithm, 
the number of iterations to evaluate the r.h.s. of Eq.~(\ref{pL.pLXup.FFT})
except the momentum space degrees of freedom becomes 
\begin{equation}
1+N_{c}^{2}+N_{c}^{2}N_{\alpha}^{2}+N_{c}^{2}N_{\alpha}^{2}+
N_{c}^{2}N_{\alpha}+N_{c}^{2}N_{\alpha}=370, 
\end{equation}
which is 
remarkably 
smaller than the numbers in naive counting when computing the 4-pt correlator, 
\begin{equation}
(N_{c}!N_{\alpha})^{B}\times 
N_{u}!N_{d}!N_{s}!\times 
2^{N_{\Lambda}+N_{\Sigma^{0}}-B} = 3456
\end{equation}
and 
$(N_{c}! N_{\alpha})^{2B} \times N_{u}! N_{d}! N_{s}! = 3 ~ 981 ~ 312$, 
where $(N_{\Lambda}, N_{\Sigma^{0}}, N_{u}, N_{d}, N_{s}, B)=(1,0,3,2,1,2)$, 
the numbers of $\Lambda$, $\Sigma^{0}$, up-quark, down-quark, 
strange-quark and the baryons, respectively. 
%
%

%
\subsection{Classification with respect to quark lines on the source side
}

Since this algorithm does not impose any restrictions on the quark fields 
on each baryon in the source, there is no need for each quark field 
in the source to be spatially identical between the baryons. 
The Wick contraction can be performed appropriately no matter what 
quantum state is considered.
This algorithm fully preserves the internal degrees of freedom 
of each three-quark 
field 
contained in the two baryons 
on the source side. 
Therefore, allowing us to classify in detail which baryon on
the source side the quark propagation in each baryon block comes from.

In the Fig.~\ref{Fig_FFT_Lp}, 
each baryon block contains three quark lines connected from the source side. 
We classify which of the two baryons on the source side those quark lines come from; 
there are $2^3=8$ possibilities in general for a baryon block. 
If all three quark lines come from the first (second) baryon, i.e., 
$\overline{p}_{\alpha_{3}}$ ($\overline{\Lambda}_{\alpha_{4}}$), 
it is labeled as [111] ([222]). 
When the three quark lines come from different baryons, 
they are indicated by one of the remained six combinations as 
[211], [121], [221], [112], [212], or [122], so as to correctly represent 
the quark lines connect from the which of the two baryons on the source side. 
%
%
\begin{table}[htbp]
\centering \leavevmode
\begin{tabular}{cccccc}
\hline
$p^{(1)},\Lambda^{(1)}$ & $p^{(2)},\Lambda^{(2)}$ & $p^{(3)},\Lambda^{(3)}$ & 
$p^{(4)},\Lambda^{(4)}$ & $p^{(5)},\Lambda^{(5)}$ & $p^{(6)},\Lambda^{(6)}$ \\
\hline
[111],[222] & [112],[221] & [121],[122] & 
[122],[121] & [112],[221] & [122],[121] \\
\hline
\end{tabular}
\caption{Classification which of two baryons on the source side the 
quark lines come from in each baryon block for the 
$\langle p\Lambda\overline{p\Lambda_{\mathrm{dsu}}}\rangle$ 
correlation function according to 
Eqs.~(\ref{pL.pLXup.FFT})~$-$~(\ref{EffectiveBaryonBlocksProtonLambdaStructuredForm}). 
The first (second) square brackets %
show the first (second) baryon block, 
i.e., proton ($\Lambda$). 
\label{Classification_of_LNdsu}}
\end{table}
%
Table~\ref{Classification_of_LNdsu} shows the classification of 
baryon blocks to calculate the 
$\langle p\Lambda \overline{p\Lambda_{\mathrm{dsu}}}\rangle$ 
correlation function 
according to 
Eqs.~(\ref{pL.pLXup.FFT})~$-$~(\ref{EffectiveBaryonBlocksProtonLambdaStructuredForm}). 
%
If we calculate only one particular single channel correlation function 
such a classification might not be very useful. %
However, in order to execute efficiently a large number of high performance 
computing jobs with huge electric power, it is beneficial to 
perform a simultaneous HPC job %
for various $BB$ channels. 
Here, for example, we will calculate the correlation functions of 52 
channels, from $NN$ to $\Xi\Xi$: 
{%
\begin{eqnarray}
&&
\!\!\!\!
\!\!\!\!\!\!\!\!
\langle pn\overline{pn}\rangle,~ 
\label{GeneralBB_NN}
\\
&&
\!\!\!\!\!\!\!\!\!\!\!\!\!\!\!\!
\begin{array}{lll}
\langle p\Lambda\overline{p\Lambda}\rangle, &  \langle p\Lambda\overline{\Sigma^{+}n}\rangle, &  \langle p\Lambda\overline{\Sigma^{0}p}\rangle, 
\\
\langle \Sigma^{+}n\overline{p\Lambda}\rangle, &  \langle \Sigma^{+}n\overline{\Sigma^{+}n}\rangle, &  \langle \Sigma^{+}n\overline{\Sigma^{0}p}\rangle, 
\\
\langle \Sigma^{0}p\overline{p\Lambda}\rangle, &  \langle \Sigma^{0}p\overline{\Sigma^{+}n}\rangle, &  \langle \Sigma^{0}p\overline{\Sigma^{0}p}\rangle, 
\end{array}
\label{GeneralBB_NL}
\end{eqnarray}
\begin{eqnarray}
&&
\!\!\!\!\!\!\!\!\!\!\!\!\!\!\!\!
\begin{array}{llllll}
\langle \Lambda\Lambda\overline{\Lambda\Lambda}\rangle, &\!\!\!\! \langle \Lambda\Lambda\overline{p\Xi^{-}}\rangle, &\!\!\!\! \langle \Lambda\Lambda\overline{n\Xi^{0}}\rangle, &\!\!\!\! \langle \Lambda\Lambda\overline{\Sigma^{+}\Sigma^{-}}\rangle, &\!\!\!\! \langle \Lambda\Lambda\overline{\Sigma^{0}\Sigma^{0}}\rangle, 
\\
\langle p\Xi^{-}\overline{\Lambda\Lambda}\rangle, &\!\!\!\! \langle p\Xi^{-}\overline{p\Xi^{-}}\rangle, &\!\!\!\! \langle p\Xi^{-}\overline{n\Xi^{0}}\rangle, &\!\!\!\! \langle p\Xi^{-}\overline{\Sigma^{+}\Sigma^{-}}\rangle, &\!\!\!\! \langle p\Xi^{-}\overline{\Sigma^{0}\Sigma^{0}}\rangle, &\!\!\!\! \langle p\Xi^{-}\overline{\Sigma^{0}\Lambda}\rangle,~
\\
\langle n\Xi^{0}\overline{\Lambda\Lambda}\rangle, &\!\!\!\! \langle n\Xi^{0}\overline{p\Xi^{-}}\rangle, &\!\!\!\! \langle n\Xi^{0}\overline{n\Xi^{0}}\rangle, &\!\!\!\! \langle n\Xi^{0}\overline{\Sigma^{+}\Sigma^{-}}\rangle, &\!\!\!\! \langle n\Xi^{0}\overline{\Sigma^{0}\Sigma^{0}}\rangle, &\!\!\!\! \langle n\Xi^{0}\overline{\Sigma^{0}\Lambda}\rangle,~
\\
\langle \Sigma^{+}\Sigma^{-}\overline{\Lambda\Lambda}\rangle, &\!\!\!\! \langle \Sigma^{+}\Sigma^{-}\overline{p\Xi^{-}}\rangle, &\!\!\!\! \langle \Sigma^{+}\Sigma^{-}\overline{n\Xi^{0}}\rangle, &\!\!\!\! \langle \Sigma^{+}\Sigma^{-}\overline{\Sigma^{+}\Sigma^{-}}\rangle, &\!\!\!\! \langle \Sigma^{+}\Sigma^{-}\overline{\Sigma^{0}\Sigma^{0}}\rangle, &\!\!\!\! \langle \Sigma^{+}\Sigma^{-}\overline{\Sigma^{0}\Lambda}\rangle,~
\\
\langle \Sigma^{0}\Sigma^{0}\overline{\Lambda\Lambda}\rangle, &\!\!\!\! \langle \Sigma^{0}\Sigma^{0}\overline{p\Xi^{-}}\rangle, &\!\!\!\! \langle \Sigma^{0}\Sigma^{0}\overline{n\Xi^{0}}\rangle, &\!\!\!\! \langle \Sigma^{0}\Sigma^{0}\overline{\Sigma^{+}\Sigma^{-}}\rangle, &\!\!\!\! \langle \Sigma^{0}\Sigma^{0}\overline{\Sigma^{0}\Sigma^{0}}\rangle,~
\\
&\!\!\!\! \langle \Sigma^{0}\Lambda\overline{p\Xi^{-}}\rangle, &\!\!\!\! \langle \Sigma^{0}\Lambda\overline{n\Xi^{0}}\rangle, &\!\!\!\! \langle \Sigma^{0}\Lambda\overline{\Sigma^{+}\Sigma^{-}}\rangle, & &\!\!\!\! \langle \Sigma^{0}\Lambda\overline{\Sigma^{0}\Lambda}\rangle, 
\end{array}
\label{GeneralBB_LL}
\\
&&
\!\!\!\!\!\!\!\!\!\!\!\!\!\!\!\!
\begin{array}{lll}
\langle \Xi^{-}\Lambda\overline{\Xi^{-}\Lambda}\rangle, & \langle \Xi^{-}\Lambda\overline{\Sigma^{-}\Xi^{0}}\rangle, & \langle \Xi^{-}\Lambda\overline{\Sigma^{0}\Xi^{-}}\rangle,~
\\
\langle \Sigma^{-}\Xi^{0}\overline{\Xi^{-}\Lambda}\rangle, & \langle \Sigma^{-}\Xi^{0}\overline{\Sigma^{-}\Xi^{0}}\rangle, & \langle \Sigma^{-}\Xi^{0}\overline{\Sigma^{0}\Xi^{-}}\rangle,~
\\
\langle \Sigma^{0}\Xi^{-}\overline{\Xi^{-}\Lambda}\rangle, & \langle \Sigma^{0}\Xi^{-}\overline{\Sigma^{-}\Xi^{0}}\rangle, & \langle \Sigma^{0}\Xi^{-}\overline{\Sigma^{0}\Xi^{-}}\rangle,~
\end{array}
\label{GeneralBB_XL}
\\
&&
\!\!\!\!
\!\!\!\!\!\!\!\!
\langle \Xi^{-}\Xi^{0}\overline{\Xi^{-}\Xi^{0}}\rangle.~
\label{GeneralBB_XX}
\end{eqnarray}
}
%
In the circumstance, %
we need to know exactly in which channel each baryon block is required, 
and which baryon on the source side each quark line is associated with 
as a result of the Wick's contraction.
%
\begin{table}[htbp]
\centering \leavevmode
\begin{tabular}{ccccccccccc}
\hline
\multicolumn{2}{c}{Baryon block}              & [111] & [211] & [121] & [221] & [112] & [212] & [122] & [222] & Total \\
\hline
Proton                   & $X_{\mathrm{udu}}$ &    18 &     0 &    31 &     0 &   106 &    16 &   121 &    12 &    304 \\
$\Sigma^{+}$             & $X_{\mathrm{usu}}$ &     3 &     0 &    10 &     0 &    52 &     3 &    55 &     1 &    124 \\
$\Xi^{0}$                & $X_{\mathrm{uss}}$ &    16 &    19 &     0 &     0 &   118 &   102 &    29 &    14 &    298 \\
\hline
$\Lambda_{\mathrm{dsu}}$ & $X_{\mathrm{dsu}}$ &   242 &   318 &   436 &   408 &   290 &   266 &   376 &   248 &   2584 \\
$\Lambda_{\mathrm{sud}}$ & $X_{\mathrm{sud}}$ &    94 &   164 &   102 &   132 &   130 &   164 &   102 &    96 &    984 \\
$\Lambda_{\mathrm{uds}}$ & $X_{\mathrm{uds}}$ &    94 &   102 &   130 &   102 &   164 &   132 &   164 &    96 &    984 \\
\hline
\end{tabular}
\caption{
Numbers how many times each baryon block is declared in the whole 
calculation of 52 correlation functions given in 
Eqs.~(\ref{GeneralBB_NN})$-$(\ref{GeneralBB_XX}). 
The baryon block is classified into eight $(=2^{3})$ forms 
with respect to 
which of 
two baryons on the source side the quark lines come from. 
\label{Classification_of_whole52channels}}
\end{table}
Table~\ref{Classification_of_whole52channels} summarizes the classification of the numbers of 
baryon blocks needed to calculate all 52 4pt-correlators %
of the 
$BB$ %
channels with respect to the forms that their quark lines 
connect. 
In the 2+1 flavor calculation, since isospin symmetry holds, 
the propagators of the up and down quarks are numerically the same,
and the classification with respect to neutron can be included in the proton. 
Similarly, the classification for $\Sigma^-$ ($\Xi^-$) is contained in the $\Sigma^+$ ($\Xi^0$). 
The classification concerning $\Sigma^0$ can be included in $\Lambda$. 
If we use the interpolating field of $\Sigma^{0}$ as 
$\Sigma^{0} = {1\over{\sqrt{2}}} \left( X_{\mathrm{dsu}} + X_{\mathrm{usd}} \right)$, 
both of the classifications for $\Sigma^{0}_{\mathrm{dsu}}$ and $\Sigma^{0}_{\mathrm{usd}}$ 
are included in $\Lambda_{\mathrm{dsu}}$. 
The numbers of $\Lambda_{\mathrm{dsu}}$ block declared in the whole calculation 
become %
larger than the numbers of other baryon blocks declared\footnote{
One can take a different expression for the interpolating field of $\Sigma^{0}$, such as 
$\Sigma^{0} = {1\over{\sqrt{2}}} \left( X_{\mathrm{dsu}} - X_{\mathrm{sud}} \right)$. 
In this case, the classification w.r.t. $\Sigma^{0}_{\mathrm{dsu}}$ ($\Sigma^{0}_{\mathrm{sud}}$) 
                         is included in $\Lambda_{\mathrm{dsu}}$    ($\Lambda_{\mathrm{sud}}$). 
The numbers for $\Lambda_{\mathrm{dsu}}$ and $\Lambda_{\mathrm{sud}}$ in the 
Table~\ref{Classification_of_whole52channels} change 
as follows ($\Lambda_{\mathrm{uds}}$ remains unchanged): 
$$
\begin{array}{cccccccccc}
\hline
                       & [111] & [211] & [121] & [221] & [112] & [212] & [122] & [222] & \mathrm{Total} \\
\hline
\Lambda_{\mathrm{dsu}} &   168 &   236 &   300 &   300 &   184 &   184 &   240 &   172 &   1784 \\
\Lambda_{\mathrm{sud}} &   168 &   300 &   184 &   240 &   236 &   300 &   184 &   172 &   1784 \\
\Lambda_{\mathrm{uds}} &    94 &   102 &   130 &   102 &   164 &   132 &   164 &    96 &    984 \\
\hline 
\end{array}
$$
} in the Table~\ref{Classification_of_whole52channels}. 

\section{Numerical results}

In this section, we present the results of the system with strangeness $S=-1$ 
at nearly physical quark masses corresponding to $(m_{\pi}, m_{K}) = (146, 525)$~MeV 
with large volume $(L a)^{4} = (96 a)^{4} = (8.1~\mathrm{fm})^{4}$. 
Earlier results had already been reported at LATTICE 2017~\cite{Nemura:2017vjc}; 
the lattice setup is basically the same as that in Ref.~\cite{Nemura:2017vjc}. 
We increase the number of statistics to the double from the 
number in the Ref.~\cite{Nemura:2017vjc}. 
For a more detail of the lattice setup, 
please refer to the earlier report~\cite{Nemura:2017vjc}.

\subsection{$\Lambda N$ potential}

In order to obtain the single-channel $\Lambda N$ potential, 
we first extract the $R$-correlator 
projected on to appropriate angular momentum $(J,M)$ state 
from the 4pt-correlator
\begin{equation}
 \begin{array}{c}
{R}_{\alpha_{1}\alpha_{2}}^{\langle p \Lambda\overline{p \Lambda}\rangle}(\vec{r},t-t_0;JM) 
= {\rm e}^{(m_{p}+m_{\Lambda})(t-t_0)} 
 \sum_{\alpha_{3}\alpha_{4}} P_{\alpha_{3}\alpha_{4}}^{(J,M)} 
 {F}_{\alpha_{1}\alpha_{2},\alpha_{3}\alpha_{4}}^{\langle p\Lambda\overline{p\Lambda}\rangle}(\vec{r},t-t_0). 
 \end{array}
\end{equation}
For the spin triplet state, the $R$ %
is further decomposed into 
the $S$- 
and 
$D$-wave components as 
\begin{equation}
 \left\{
 \begin{array}{l}
    R(\vec{r};\ ^3S_1)={\cal P}R(\vec{r};J=1)\equiv 
    {1\over 24} \sum_{{\cal R}\in{ O}} {\cal R}R(\vec{r};J=1),
   \\
  R(\vec{r};\ ^3D_1)={\cal Q}R(\vec{r};J=1)\equiv 
  (1-{\cal P})R(\vec{r};J=1).
  \end{array}
 \right.
\end{equation}
Two central and tensor potentials, 
$V^{\mathrm{(Central)}}(r;J=0)=(V^{(0)}(r)-3V^{(\sigma)}(r))$ for $J=0$, 
$V^{\mathrm{(Central)}}(r;J=1)=(V^{(0)}(r) +V^{(\sigma)}(r))$, 
and $V^{\mathrm{(Tensor)}}(r)$ for $J=1$, 
are determined from the Schr\"{o}dinger equation. 
\begin{equation}
 \left\{
 \begin{array}{c}
  V^{\mathrm{(C)}}(r;J=0){ R}(\vec{r},t-t_0;J=0)
  =\left({\nabla^2\over 2\mu}-{\partial\over \partial t}\right){ R}(\vec{r},t-t_0;J=0), 
\\
 \left\{
 \begin{array}{c}
  {\cal P} \\
  {\cal Q}
 \end{array}
 \right\}
 \times
 \left\{
  V^{\mathrm{(C)}}(r;J=1)+V^{(T)}(r)S_{12}
 \right\}
 { R}(\vec{r},t-t_0;J=1)

 =
 \left\{
 \begin{array}{c}
  {\cal P} \\
  {\cal Q}
 \end{array}
 \right\}
 \times
 \left\{
     {\nabla^2\over 2\mu%
                                  } 
     -{\partial \over \partial t}
 \right\}
 { R}(\vec{r},t-t_0;J=1).
 \end{array}
 \right.
\end{equation}
%
%

%
\begin{figure}[t]
 \begin{minipage}[t]{0.33\textwidth}
  \centering \leavevmode
%
  \includegraphics[width=0.99\textwidth]{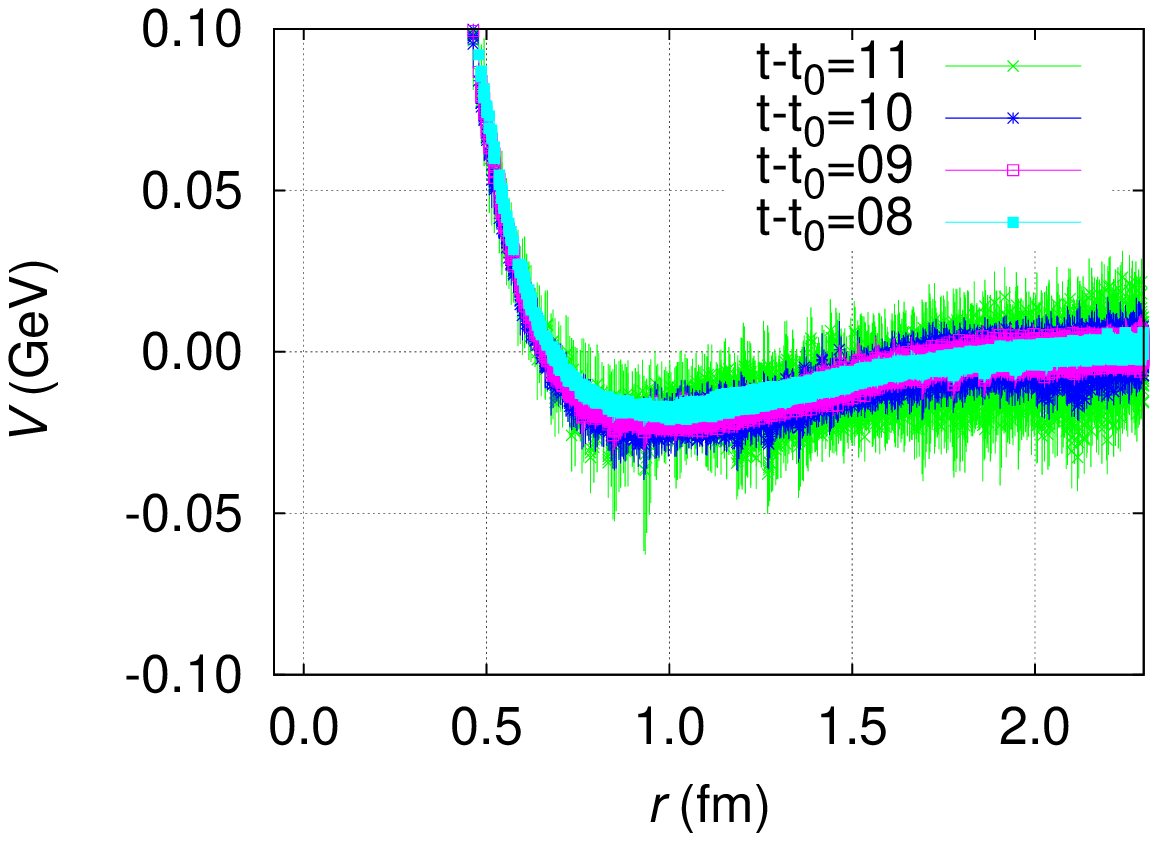}
%
 \end{minipage}~
 \hfill
 \begin{minipage}[t]{0.33\textwidth}
  \centering \leavevmode
%
  \includegraphics[width=0.99\textwidth]{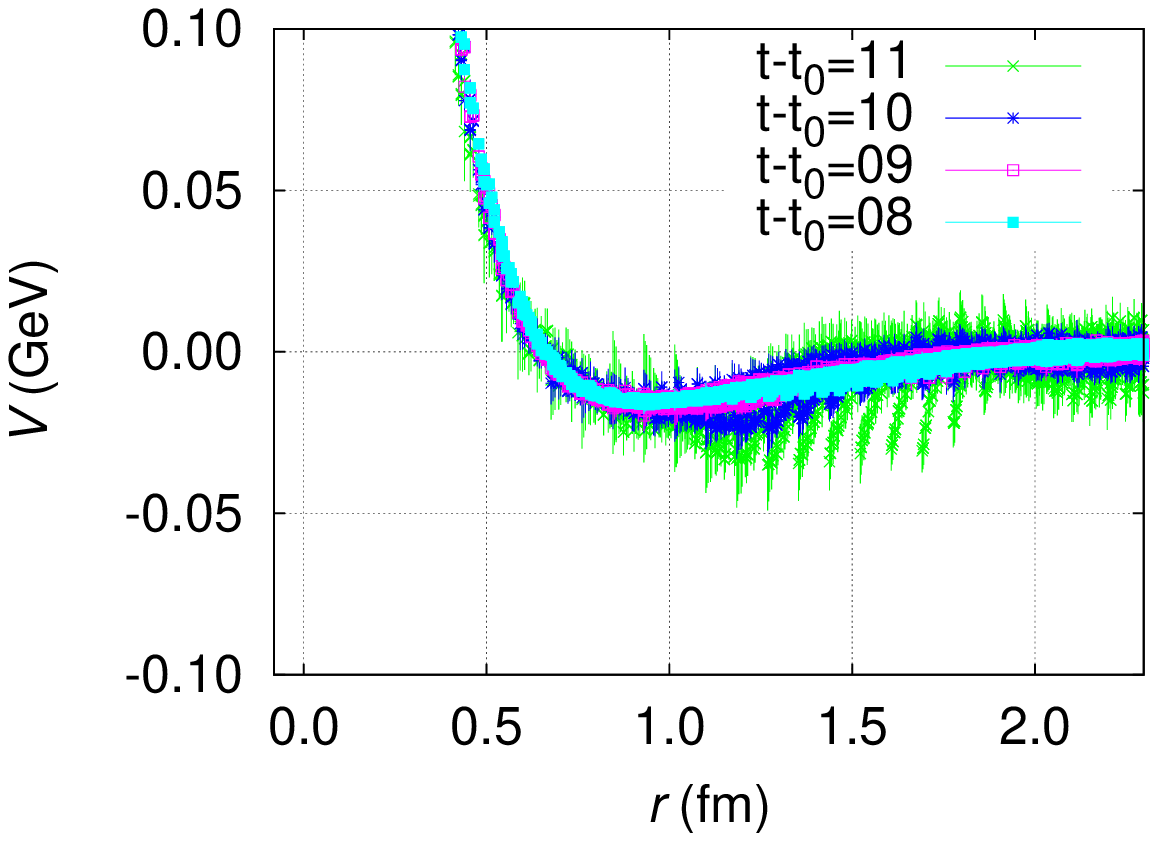}
%
 \end{minipage}~
 \hfill
 \begin{minipage}[t]{0.33\textwidth}
  \centering \leavevmode
%
  \includegraphics[width=0.99\textwidth]{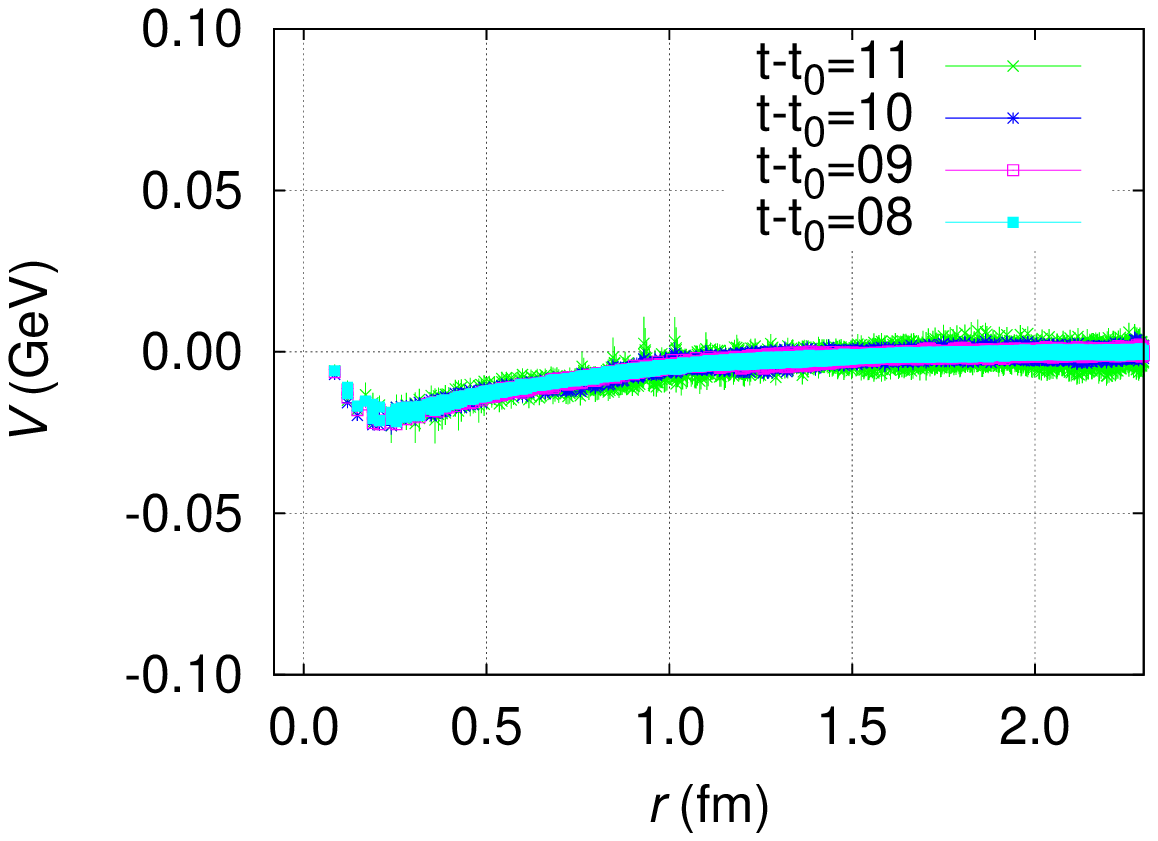}
%
 \end{minipage}
 \bigskip
 \caption{Three $\Lambda N$ single-channel potentials of 
   (i) $^1S_0$ central (left), 
   (ii) $^3S_1-^3D_1$ central (center), and 
   (iii) $^3S_1-^3D_1$ tensor (right). 
   \label{VCeff1S0_VC3E1_VT3E1_LN_from_1x1}}
\end{figure}
%
%

\begin{figure}[t]
 \begin{minipage}[t]{0.33\textwidth}
  \centering \leavevmode
%
  \includegraphics[width=0.99\textwidth]{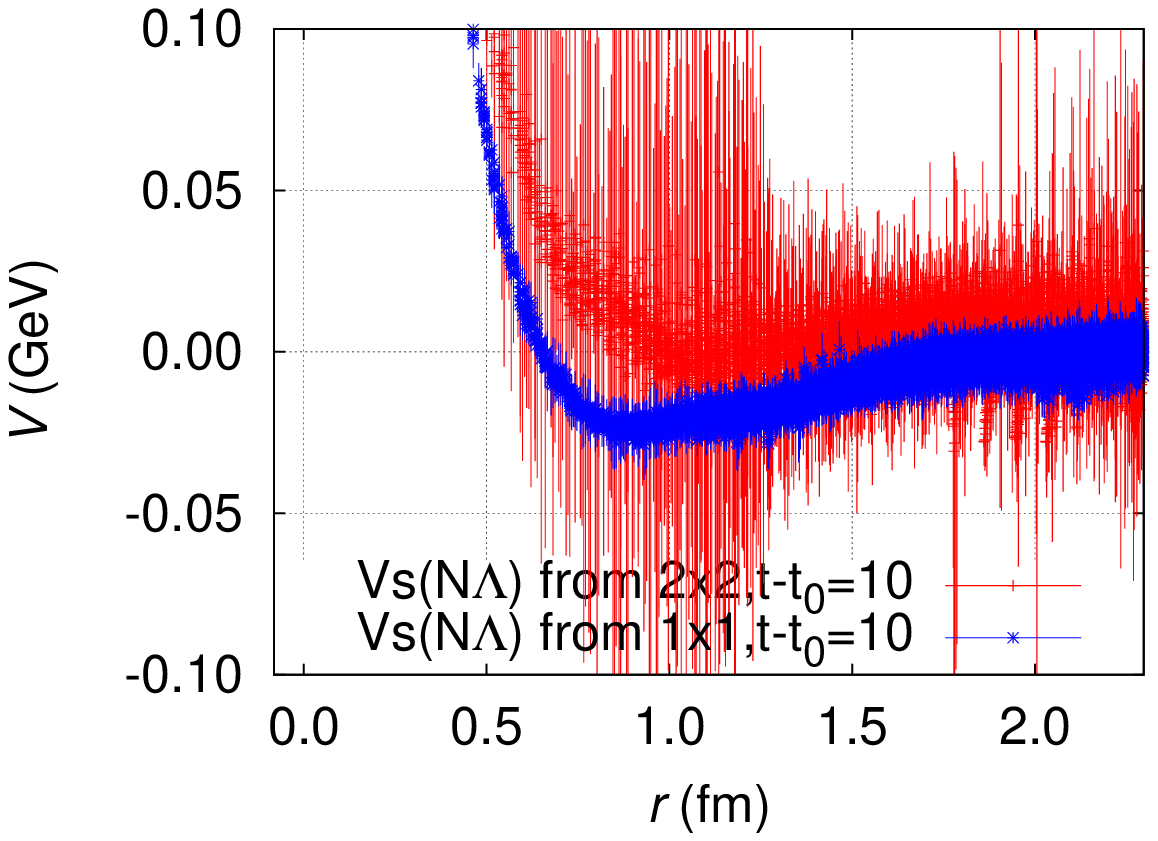}
%
 \end{minipage}~
 \hfill
 \begin{minipage}[t]{0.33\textwidth}
  \centering \leavevmode
%
  \includegraphics[width=0.99\textwidth]{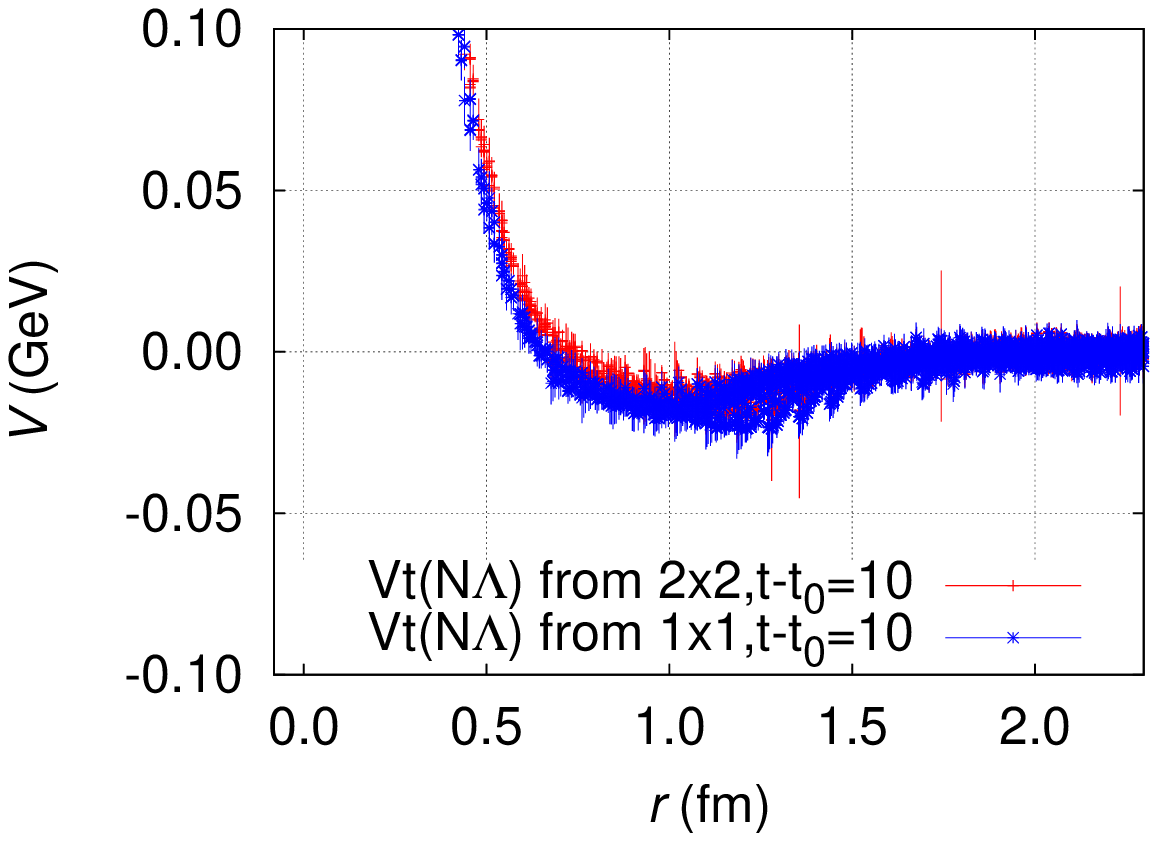}
%
 \end{minipage}~
 \hfill
 \begin{minipage}[t]{0.33\textwidth}
  \centering \leavevmode
%
  \includegraphics[width=0.99\textwidth]{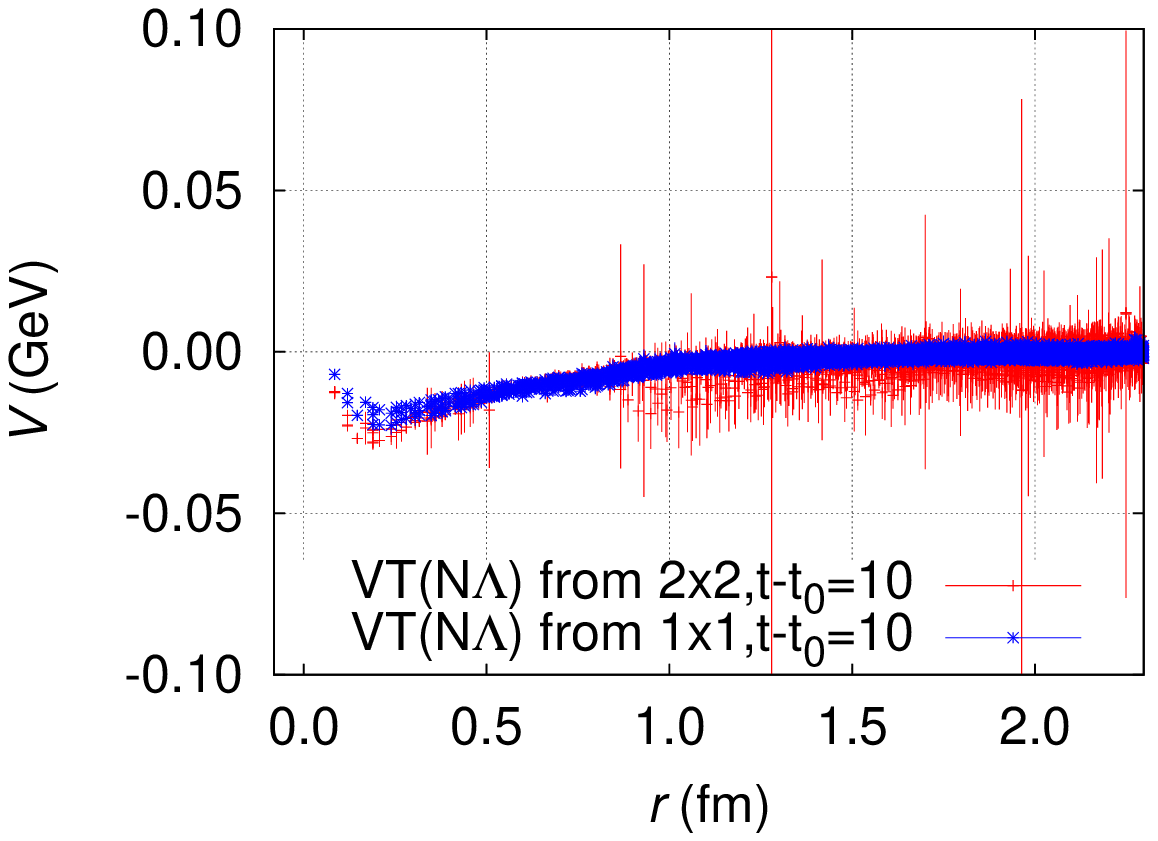}
%
 \end{minipage}
 \bigskip
 \caption{Comparisons between the $\Lambda N-\Lambda N$ diagonal part in $2\times 2$ coupled-channel potential and 
the $\Lambda N$ single channel potential, for 
   (i) $^1S_0$ central (left), 
   (ii) $^3S_1-^3D_1$ central (center), and 
   (iii) $^3S_1-^3D_1$ tensor (right). 
   \label{Compare_VCeff1S0_VC3E1_VT3E1_LN_from_2x2_vs_from_1x1}}
\end{figure}
%
%
Fig.~\ref{VCeff1S0_VC3E1_VT3E1_LN_from_1x1} shows 
three potentials of the $\Lambda N$ system; 
(i)~the central potential in the $^1S_0$ (left), 
(ii)~the central potential in the $^3S_1-^3D_1$ (center), and 
(iii)~the tensor  potential in the $^3S_1-^3D_1$ (right). %
These potentials are obtained through the single channel formulation 
of the HAL QCD method, which are valid below the $\Sigma N$ threshold 
and the effects by coupling with the $\Sigma N$ channel are implicitly 
included. 
Both the $^1S_0$ and $^3S_1-^3D_1$ central potentials have 
short ranged repulsive core and medium-to-long-distanced 
attractive well. 
These two potentials are more or less similar to each other. 
For flavor $SU(3)$ symmetric case, 
the $\Lambda N$ state is represented in terms of 
irreducible representation ({\it irrep.}) in flavor space as 
$
|\Lambda N,^1S_0\rangle = {1\over{\sqrt{10}}} 
\left( 3|{\bm{27}}\rangle + |{\bm{8}}_{\mathrm{s}}\rangle\rangle \right)$, and 
$
|\Lambda N,^3E_1\rangle = {1\over{\sqrt{2}}} 
\left( |\overline{\bm{10}}\rangle - |{\bm{8}}_{\mathrm{a}}\rangle \right)$, 
while the $NN$ state is given by 
$
|NN,~^1S_0\rangle = |{\bm{27}}\rangle$, and %
$
|NN,~^3E_1\rangle = |\overline{\bm{10}}\rangle$. 
These single channel $\Lambda N$ $^1S_0$ and $^3S_1-^3D_1$ potentials are 
qualitatively %
similar to the $NN$ $^1S_0$ and $^3S_1-^3D_1$ potentials, respectively, 
though the flavor structures of the $\Lambda N$ are not the same 
as those of the $NN$. 
Observing the central potentials in detail, 
the $^1S_0$ potential gradually changes to become a little attractive 
when the time slice increases from $t=8$ to $9$ but the change is 
unclear at $t=10,11$ due to the large statistical uncertainty; 
this behavior reflects to the scattering phase shifts shown in the 
next subsection. 
On the other hand, such a deviation of the potential in the 
$^3S_1-^3D_1$ channel is little; the central values of the potential 
hardly change over time to time but the statistical uncertainty increases. 

Fig.~\ref{Compare_VCeff1S0_VC3E1_VT3E1_LN_from_2x2_vs_from_1x1} 
shows the comparisons between the $\Lambda N$ potential 
from the single channel analysis 
and 
the $\Lambda N-\Lambda N$ diagonal part 
from the coupled-channel analysis 
using $\Lambda N$ and $\Sigma N$ correlation functions. 
The statistical uncertainty of the $\Lambda N-\Lambda N$ 
diagonal part of the coupled-channel potential 
in the $^1S_0$ is large and it 
dramatically reduces 
by taking the single channel $\Lambda N$ potential. 
For the $^3S_1-^3D_1$ state, the single channel $\Lambda N$ central 
potential slightly enhances the attraction from the $\Lambda N-\Lambda N$ 
diagonal part of %
the coupled-channel potential. 
The short distance part of the single channel $\Lambda N$ tensor 
potential is slightly suppressed from the $\Lambda N-\Lambda N$ 
diagonal part of %
the coupled-channel potential.

\subsection{Scattering phase shifts}

%
%
%
%
\begin{wrapfigure}[8]{R}[0pt]{0.40\textwidth}
  \vspace*{-32pt}
  \hspace*{32pt}
  \includegraphics[width=0.3267\textwidth]{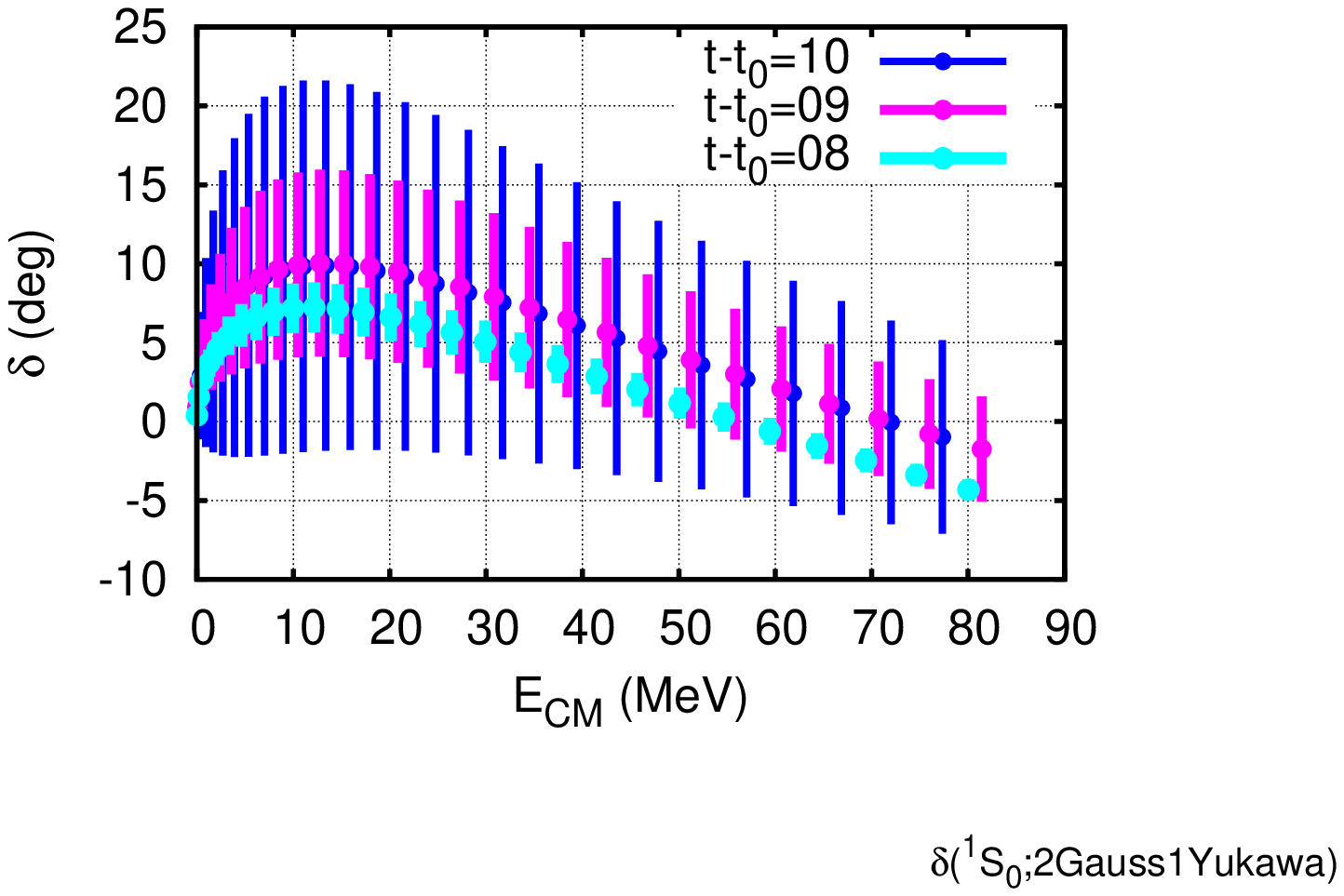}
\\
 \caption{Scattering phase shift in the $^1S_0$ state 
   of $\Lambda N$ system, 
   obtained by 
   solving the Schr\"{o}dinger equation with %
   parametrized functional form in Ref.~\cite{Nemura:2017vjc}. %
   \label{Fig_Phsft1S0LN}}
\end{wrapfigure}
%
%
%
%
By employing the single channel analysis for the $\Lambda N$ low-energy 
state %
the statistical errors are 
significantly reduced. 
Thus we parametrize the potentials %
with the %
analytic functional form in Ref.~\cite{Nemura:2017vjc}. 
Figure~\ref{Fig_Phsft1S0LN} shows the scattering phase shift in $^1S_0$ 
channel %
obtained through 
the %
parametrized $\Lambda N$ potential. 
The present result shows that the interaction in the $^1S_0$ channel 
is attractive on average in the low-energy region 
though the fluctuation is large. 

%
\begin{figure}[b]
 \begin{minipage}[t]{0.33\textwidth}
  \centering \leavevmode
  \includegraphics[width=0.99\textwidth]{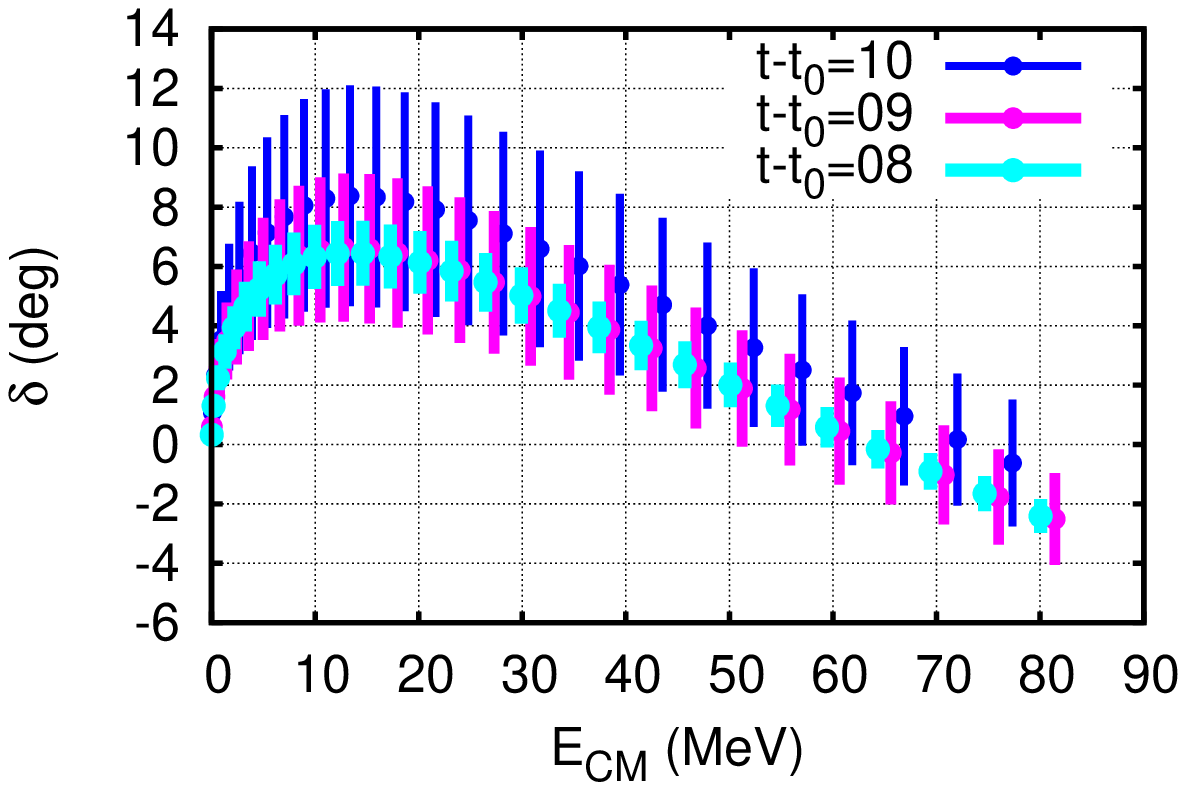}
 \end{minipage}~
 \hfill
 \begin{minipage}[t]{0.33\textwidth}
  \centering \leavevmode
  \includegraphics[width=0.99\textwidth]{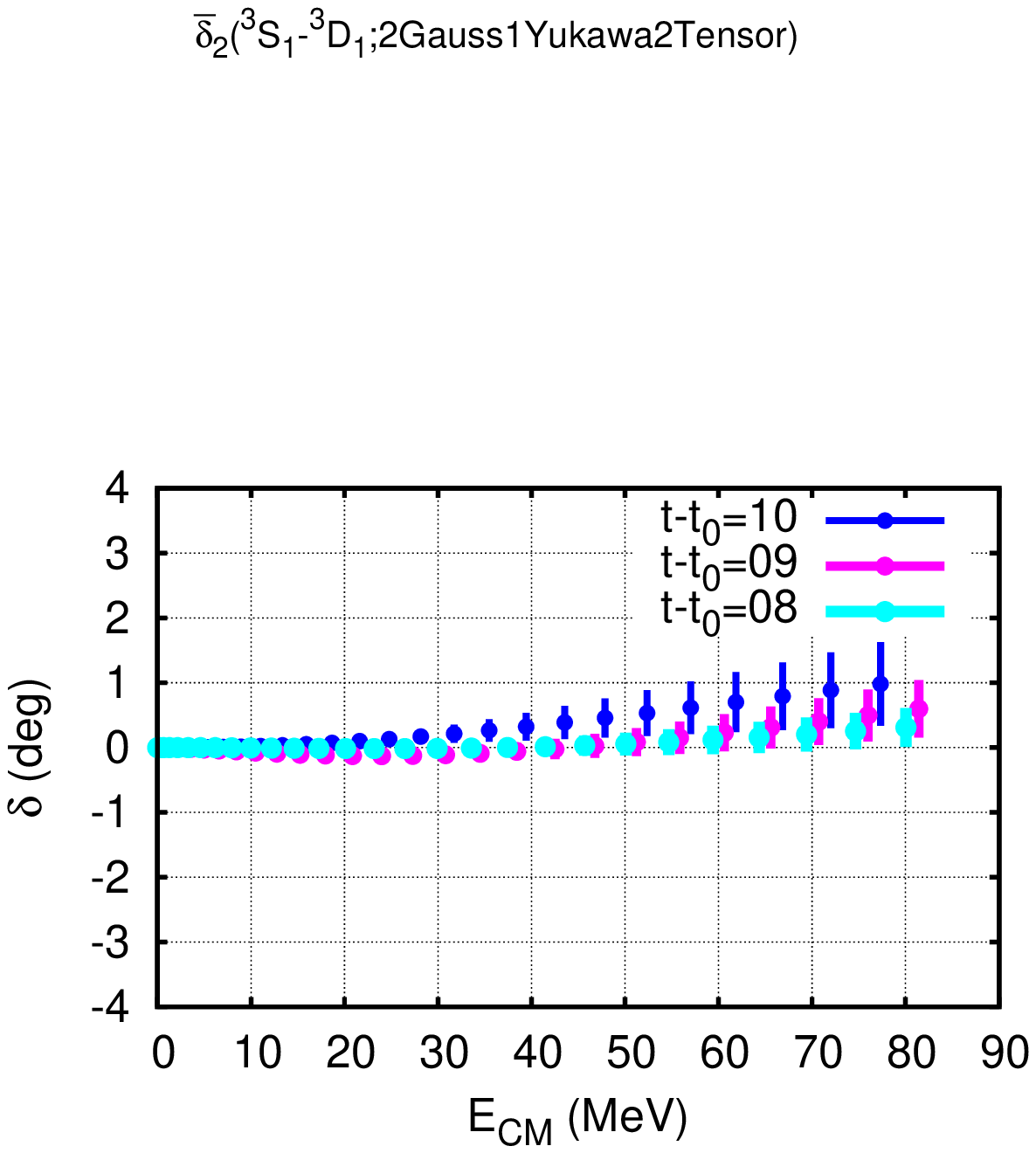}
 \end{minipage}~
 \hfill
 \begin{minipage}[t]{0.33\textwidth}
  \centering \leavevmode
  \includegraphics[width=0.99\textwidth]{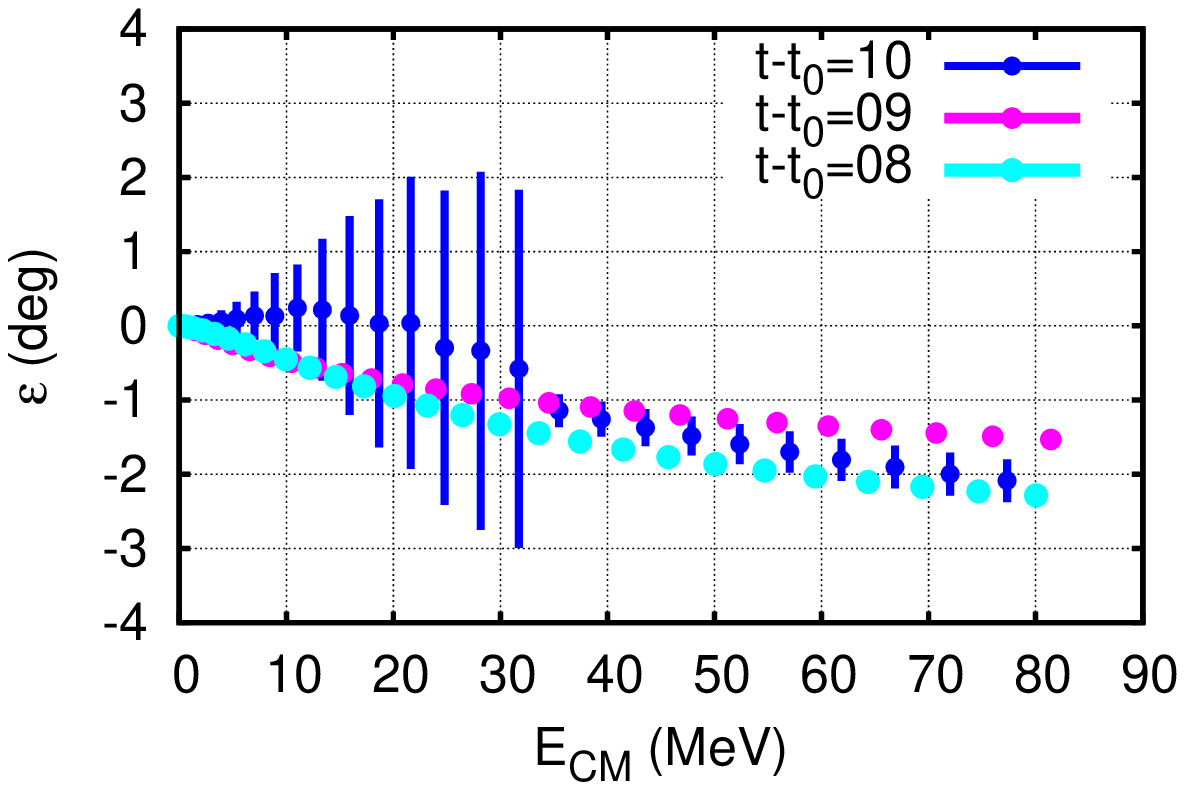}
 \end{minipage}
 \bigskip
  \caption{Scattering bar-phase shifts and mixing angle in the 
    $^3S_1-^3D_1$ states of $\Lambda N$ system, 
    $\bar{\delta}_0$ (left), 
    $\bar{\delta}_2$ (center), and 
    $\bar{\varepsilon}_1$ (right), 
    obtained by solving the Schr\"{o}dinger equation 
    with parametrized functional form in Ref.~\cite{Nemura:2017vjc}. %
    \label{Fig_Phsft3E1LN}}
\end{figure}
%
%
%
Figure~\ref{Fig_Phsft3E1LN} shows the scattering phase shifts in 
$^3S_1-^3D_1$ channels. 
For the $^3S_1-^3D_1$ channels, 
the scattering matrix is parametrized with three real parameters 
bar-phase shifts and mixing angle. %
The phase shift $\bar{\delta}_0$ at the time slices $t-t_0=8-10$ shows 
the interaction is attractive. 
Because of large statistical uncertainty, 
it is hard to see how large is the spin-spin interaction $(V^{(\sigma)})$ 
of the $\Lambda N$ interaction. 

Both the strengths of attraction in $\delta(^1S_0)$ and $\bar{\delta}_{0}$ 
seem to be weaker than the usual empirical values such as 
Ref.~\cite{Haidenbauer:2019boi}; it might be due to a little deviation 
of light quark mass which is corresponding to 
$(m_{\pi},m_{K})\approx(146,525)$~MeV. 
For both Figs.~\ref{Fig_Phsft1S0LN} and \ref{Fig_Phsft3E1LN}, the parametrizing procedure is not very stable, 
especially for $V^{(\sigma)}$ and $V^{(T)}$. 
The present phase shifts and mixing angle are preliminary results at this moment.

\section{Discussion
}

We present the phase shifts of low energy $\Lambda N$ scattering 
by using the potential extracted from lattice QCD through the HAL QCD method. 
Both the spin singlet and triplet states are weakly attractive in low energy region, 
which are qualitatively in good agreement with empirical studies.
However, the strengths of the attraction seem to be weaker than the phenomenological conclusion. 
In addition, statistical uncertainty is still large to pin down the spin-spin interaction of the $\Lambda N$ system. 
Since the present results still have a large uncertainty, we need to continue our calculations based on the following points.
(i) employ the physical quark masses which is more precisely close to the real system 
(but practically we should apply the 2+1 flavor approach for a while), 
(ii) using a sufficiently large volume, 
(iii) to endeavor for clarifying various details to reduce the statistical and/or systematic uncertainty. 

In this report, the numerical results are obtained by using the wall quark sources, where 
the quark field is uniform and identical in all spatial directions even between the baryons. 
The effective block algorithm presented in this report can be applied for 
more flexible choices of interpolating field on the source side; 
this should be a strong advantageous to improve the present large uncertainty. 
We hope an improved calculation will be reported in the future.

\acknowledgments

We thank all collaborators in this project, above all, 
members of PACS Collaboration for the gauge configuration generation, 
and members of HAL QCD Collaboration for the correlation function measurement. 
The lattice QCD calculations have been performed 
on the K computer at RIKEN, AICS 
(hp120281, hp130023, hp140209, hp150223, hp150262, hp160211, hp170230),
HOKUSAI FX100 computer at RIKEN, Wako (
G15023, G16030, G17002)
and HA-PACS at University of Tsukuba 
(14a-25, 15a-33, 14a-20, 15a-30).
We thank ILDG/JLDG~%
which serves as an essential infrastructure in this study.
This work is supported in part by HPCI System
Research Project (hp2101065). 
This work is supported in part by 
MEXT Grant-in-Aid for Scientific Research 
(JP16K05340, JP18H05236), 
and SPIRE (Strategic Program for Innovative Research) Field 5 project and 
``Priority issue on Post-K computer'' (Elucidation of the Fundamental Laws
and Evolution of the Universe) and 
Joint Institute for Computational Fundamental Science (JICFuS).
%
%
%

%
%

\providecommand{\href}[2]{#2}\begingroup\raggedright\endgroup

\if false

\fi


\begin{thebibliography}{1}

\bibitem{Demorest:2010bx}
P.~Demorest, T.~Pennucci, S.~Ransom, M.~Roberts and J.~Hessels, \emph{{Shapiro
  Delay Measurement of A Two Solar Mass Neutron Star}},
  \href{https://doi.org/10.1038/nature09466}{\emph{Nature} {\bfseries 467}
  (2010) 1081} [\href{https://arxiv.org/abs/1010.5788}{{\ttfamily 1010.5788}}].

\bibitem{Antoniadis:2013pzd}
J.~Antoniadis et~al., \emph{{A Massive Pulsar in a Compact Relativistic
  Binary}}, \href{https://doi.org/10.1126/science.1233232}{\emph{Science}
  {\bfseries 340} (2013) 6131}
  [\href{https://arxiv.org/abs/1304.6875}{{\ttfamily 1304.6875}}].

\bibitem{Fonseca:2021wxt}
E.~Fonseca et~al., \emph{{Refined Mass and Geometric Measurements of the
  High-mass PSR J0740+6620}},
  \href{https://doi.org/10.3847/2041-8213/ac03b8}{\emph{Astrophys. J. Lett.}
  {\bfseries 915} (2021) L12}
  [\href{https://arxiv.org/abs/2104.00880}{{\ttfamily 2104.00880}}].

\bibitem{Aoki:2012tk}
{\scshape HAL QCD} collaboration, \emph{{Lattice QCD approach to Nuclear
  Physics}}, \href{https://doi.org/10.1093/ptep/pts010}{\emph{PTEP} {\bfseries
  2012} (2012) 01A105} [\href{https://arxiv.org/abs/1206.5088}{{\ttfamily
  1206.5088}}].

\bibitem{Nemura:2014eta}
{\scshape HAL QCD} collaboration, \emph{{An implementation of hybrid parallel
  C++ code for the four-point correlation function of various baryon-baryon
  systems}}, \href{https://doi.org/10.22323/1.187.0426}{\emph{PoS} {\bfseries
  LATTICE2013} (2014) 426}.

\bibitem{Nemura:2017vjc}
H.~Nemura et~al., \emph{{Baryon interactions from lattice QCD with physical
  masses --- strangeness $S=-1$ sector ---}},
  \href{https://doi.org/10.1051/epjconf/201817505030}{\emph{EPJ Web Conf.}
  {\bfseries 175} (2018) 05030}
  [\href{https://arxiv.org/abs/1711.07003}{{\ttfamily 1711.07003}}].

\bibitem{Nemura:2015yha}
H.~Nemura, \emph{{Instructive discussion of an effective block algorithm for
  baryon–baryon correlators}},
  \href{https://doi.org/10.1016/j.cpc.2016.05.014}{\emph{Comput. Phys. Commun.}
  {\bfseries 207} (2016) 91}
  [\href{https://arxiv.org/abs/1510.00903}{{\ttfamily 1510.00903}}].

\bibitem{Nemura:2016sty}
H.~Nemura et~al., \emph{{A Fast Algorithm for Lattice Hyperonic Potentials}},
  \href{https://doi.org/10.7566/JPSCP.17.052002}{\emph{JPS Conf. Proc.}
  {\bfseries 17} (2017) 052002}
  [\href{https://arxiv.org/abs/1604.08346}{{\ttfamily 1604.08346}}].

\bibitem{Haidenbauer:2019boi}
J.~Haidenbauer, U.G.~Mei\ss{}ner and A.~Nogga,
  \emph{{Hyperon\textendash{}nucleon interaction within chiral effective field
  theory revisited}},
  \href{https://doi.org/10.1140/epja/s10050-020-00100-4}{\emph{Eur. Phys. J. A}
  {\bfseries 56} (2020) 91} [\href{https://arxiv.org/abs/1906.11681}{{\ttfamily
  1906.11681}}].

\end{thebibliography}

\begin{thebibliography}{99}
\bibitem{...}
....
\end{thebibliography}
\end{document}